\documentclass[11pt]{article}
\usepackage{hyperref}
\usepackage{amssymb}
\usepackage{graphicx}

\usepackage{float}
\usepackage{tikz}
\usepackage{slashed}
\usepackage[title]{appendix}
\usepackage{amsmath}
\usepackage{breqn}
\usepackage{authblk}
\usepackage{array}
\usepackage{dsfont}
\usepackage{cite}
\usepackage[section]{placeins}
\usepackage[a4paper, total={6.8in, 10in}]{geometry}
\numberwithin{equation}{section}
\DeclareMathOperator{\cosech}{cosech}
\newcommand{\be}{\begin{equation}}
\newcommand{\ee}{\end{equation}}
\newcommand{\bea}{\begin{eqnarray}}
\newcommand{\eea}{\end{eqnarray}}
\newcommand{\ba}{\begin{aligned}}
\newcommand{\ea}{\end{aligned}}
\begin{document}
\title{Thermal Casimir effect in $\kappa$-Minkowski space-time}

\author[1]{Suman Kumar Panja\thanks{sumanpanja19@gmail.com}}
\affil[1]{Centre of Excellence ENSEMBLE3 Sp. z o. o., Wolczynska Str. 133, 01-919, Warsaw, Poland}

\author[2]{Vishnu Rajagopal\thanks{vishnu@hunnu.edu.cn}} 
\affil[2]{Department of Physics and Synergetic Innovation\\
Center for Quantum Effects and Applications\\
Hunan Normal University, Changsha, Hunan 410081, China}

\maketitle

\begin{abstract}

We study the finite temperature Casimir effect for parallel plates in the $\kappa$-Minkowski space-time. Using the Matsubara formalism and imposing the Dirichlet boundary conditions on a massless $\kappa$-scalar field, we compute the $\kappa$-deformed corrections to thermal Casimir free energy, pressure, entropy, and internal energy. Our results demonstrate that space-time non-commutativity enhances the attractive nature of the thermal Casimir force while preserving thermodynamic consistency; the system satisfies the Nernst theorem and laws of thermodynamics remain intact in $\kappa$-deformed space-time. Our analysis yields an upper bound on the deformation parameter as $a\leq10^{-18}m$. Furthermore, our results indicate that non-commutative effects become experimentally observable in Casimir effect studies when the ratio of the non-commutative scale to plate separation satisfies $a/L\leq 10^{-12}$. We also obtain the expression for Stefan-Boltzmann's law in $\kappa$-Minkowski space-time. 

\end{abstract}

\section{Introduction} \label{intro}

The unification of the four fundamental interactions at the Planck scale remains one of the most intriguing challenges in modern physics. This unification of all four fundamental interactions requires a consistent quantum theory of gravity. The construction of a consistent quantum theory of gravity remains an active pursuit in theoretical physics. All approaches to construct quantum gravity inherently exhibit a fundamental length scale. Since models built on non-commutative space-times introduce a length scale, the relevance of such space-times to quantum gravity is being investigated vigorously \cite{CH-4-connes, CH-4-dop}.

Non-commutative (NC) space-times inherently introduce a minimal length scale, making it of intrinsic interest to study the implications of this scale in physical phenomena. One such phenomenon where the length scale plays a significant role is the Casimir effect \cite{hbg,pule}. It was shown that when two uncharged conducting plates are placed parallel in a small separation, they experience an attractive force \cite{hbg,pule,kam,b-m-m,Mil}. Modifications to the Casimir effect due to different dimensions of space-time, the presence of medium between plates, the thermal effect, etc., using real scalar field theory have been studied extensively \cite{kam,pule,b-m-m,Brevik}. High-precision experiments have quantified the Casimir force between parallel plates at micrometer-scale separations \cite{onofr,brax,sedmik}. Rigorous theoretical and experimental studies have also explored the Casimir effect, beyond parallel conducting plates, in various geometric configurations \cite{sed,Bimonte,Wang}. Casimir effect for various other configurations, such as piston configuration with imposed curved boundary conditions and in the presence of compact dimensions, has also been extensively studied \cite{oikonomou1}.

The Casimir effect has been investigated in various NC space-times, such as Moyal space-time \cite{CH-4-casadio}, $\kappa$-deformed space-time \cite{CH-4-pinto,kappa-casimir-korea,CH-4-skp,kappa-spherical,CH-4-skp3,CH-4-skp4}, Snyder space-time \cite{snyder} and Doplicher-Fredenhagen-Roberts (DFR) space-time \cite{CH-4-skp2}, respectively. This $\kappa$-deformed space-time is shown to be associated with the low-energy limit of specific quantum gravity models \cite{CH-4-jerzy} and its coordinates adhere to the following commutation relations
\begin{equation}
 [\hat{x}^i, \hat{x}^j]=0,~~ [\hat{x}^0,\hat{x}^i]=i a \hat{x}^i,~~ a\equiv\frac{1}{\kappa},\label{Chap-4-intro2}
\end{equation} 
where $a$  is the deformation parameter with the dimension of length. The symmetry algebras of NC space-times are typically described by a deformed Poincar\'e algebra. For $\kappa$-Minkowski space-time, this algebra takes the form of the $\kappa$-Poincar\'e algebra \cite{kappa-poincare}, which reduces to the standard Poincar\'e algebra in the corresponding commutative limits. However, \cite{CH-4-hopf,CH-4-mel2,CH-4-mel3} demonstrates that the symmetry algebra can alternatively be formulated using an undeformed $\kappa$-Poincar\'e algebra, preserving the standard Poincar\'e algebra structure while introducing deformations in the co-algebra sector. This approach has been explored using a realisation method, where NC space-time coordinates are expressed as functions of commutative coordinates and their conjugates. Furthermore, \cite{CH-4-mel3} establishes a direct correspondence between different realisation choices and ordering prescriptions. This framework simplifies the analysis of physical problems, making it the preferred method for this work.

The $\kappa$-deformed corrections to the Casimir energy between two parallel conducting plates have been calculated in \cite{CH-4-pinto,kappa-casimir-korea} by computing the zero-point energy using the $\kappa$-deformed dispersion relation, compatible with the $\kappa$-Poincar\'e algebra. In \cite{CH-4-skp}, this has been studied employing the Green's function approach to the $\kappa$-deformed scalar field theory (invariant under the undeformed $\kappa$-Poincar\'e algebra) obeying the Dirichlet boundary condition. The Casimir energy of a spherical shell in $\kappa$-Minkowski space-time for a complex scalar field, with asymmetric ordering, has been studied in \cite{kappa-spherical}. Recently, the fall of the $\kappa$-deformed Casimir energy between two parallel plates in a $\kappa$-deformed weak gravitational field and the response of the $\kappa$-deformed Casimir energy associated with parallel plates in a $\kappa$-deformed rotating frame have been studied in \cite{CH-4-skp3} and \cite{CH-4-skp4}, respectively. These studies reveal that the Casimir energy behaves analogously to the standard mass in accordance with the equivalence principle, while falling off in the weak gravitational field and responding to the centripetal forces in the $\kappa$-deformed space-time. 

Although the effects of the non-commutativity on the zero-temperature Casimir effect have been studied to some extent, a deeper understanding of how the space-time non-commutativity affects the vacuum fluctuations at a finite temperature is still an intriguing question. Moreover, the finite temperature Casimir force in DFR space-time is shown to have a non-trivial dependence on temperature and plate separation, due to the extra spatial dimensions introduced by the non-commutativity \cite{CH-4-skp2}, which necessitates the possibility of extending a similar analysis to other NC space-time models. 

In this study, we conduct a detailed investigation of the finite-temperature Casimir effect within the $\kappa$-Minkowski space-time framework. The finite-temperature Casimir effect in commutative space-times has been studied rigorously using various approaches \cite{kam,pule, Brevik,Teo-1,geyer,Mota,zhang,zhang1,godel}. The Matsubara formalism \cite{matsubara1,matsubara2} is an elegant approach to analyse the finite-temperature quantum field theories. Through Wick rotation and imposing periodic (anti-periodic) boundary conditions for bosonic (fermionic) fields, it discretises the energy spectrum into Matsubara frequencies, which enable us to evaluate the thermal expectation values using Euclidean path integrals. Here we incorporate this formalism to systematically investigate thermal fluctuations and their thermodynamic properties in $\kappa$-Minkowski space-time by constructing a $\kappa$-deformed Euclidean path integral. Using this, we investigate the finite-temperature Casimir effect for a massless $\kappa$-deformed scalar field confined between two parallel plates subjected to Dirichlet boundary conditions. We adopt the regularisation procedure discussed in \cite{zhang,zhang1}, where the renormalised thermal Casimir free energy expression is obtained by subtracting the thermal Casimir energy contributions at infinite plate separation from the formerly derived thermal Casimir energy expression. This facilitates the derivation of the corresponding analytical expressions for entropy and shows that it consistently approaches zero as the temperature vanishes, satisfying the Nernst heat theorem in the $\kappa$-Minkowski space-time. These analyses show that our results are consistent with the thermodynamic laws under the $\kappa$-deformation of space-time.

The organisation of this paper is as follows. The Sec. 2 provides a brief review on the $\kappa$-deformed space-time and its symmetry algebra. We construct the $\kappa$-deformed Lagrangian using the deformed quadratic Casimir operator. In Sec. 3, we construct the modified partition function for the $\kappa$-deformed scalar field, using the Matsubara formalism and Dirichlet boundary conditions, and then calculate the expression for the $\kappa$-deformed Casimir free energy, Casimir pressure, and entropy, along with a comprehensive analysis of these results. In Sec. 4, the $\kappa$-deformed corrections to the Stefan-Boltzmann law are obtained from the formerly constructed $\kappa$-deformed partition function. The concluding remarks and discussions of our results are summarised in Sec. 5.

\section{Lagrangian in $\kappa$-Minkowski space-time} \label{symm}

This section briefly outlines the procedure for constructing the $\kappa$-deformed scalar field Lagrangian for studying the Casimir effect at finite temperature in $\kappa$-Minkowski space-time. The deformed Lagrangian is constructed from the $\kappa$-deformed quadratic Casimir operator of the undeformed $\kappa$-Poincar\'e algebra, as discussed in \cite{CH-4-skp}. This procedure utilises the realisation approach, where the NC coordinates are defined as a function of the commutative coordinates and their derivatives \cite{CH-4-hopf,CH-4-mel2,CH-4-mel3}. Following \cite{CH-4-hopf}, we write down the $\kappa$-deformed space-time coordinates, satisfying Eq.(\ref{Chap-4-intro2}), in terms of the commutative coordinates $x_{\mu}$ and their derivatives $\partial_{\mu}$ as
\begin{equation}\label{Chap-4-CFK2}
\begin{split}
  \hat{x}_i=x_i\varphi(A),~~\hat{x}_0=x_0\psi(A)+iax_j\partial_j\gamma(A)
\end{split}
\end{equation}
where $A=ia\partial_0$. Substituting Eq.(\ref{Chap-4-CFK2}) in Eq.(\ref{Chap-4-intro2}) we obtain $\frac{d(\ln{\varphi(A)})}{dA}=\frac{\gamma(A)-1}{\psi(A)}$, together with conditions $\psi(0)=1,~\varphi(0)=1$. A consistent solution to this $\varphi(A)$ is shown by considering two possible realisations of $\psi(A)$ as $\psi(A)=1$ and $\psi(A)=1+2A$, respectively \cite{CH-4-hopf}, for any arbitrary constant $\gamma(A)$. For $\psi(A)=1$, the allowed choices for $\varphi(A)$ are obtained as $e^{-A}$, $e^{-\frac{A}{2}}$, $1$, and $\frac{A}{e^A-1}$ \cite{CH-4-hopf}. Now we carry out the upcoming calculations by choosing $\varphi=e^{-\frac{A}{2}}$. As a result, the $\kappa$-deformed space-time coordinates defined in Eq.(\ref{Chap-4-CFK2}) becomes $\hat{x}_i=x_i e^{-\frac{A}{2}}$ and $\hat{x}_0=x_0$, respectively.

In \cite{CH-4-hopf,CH-4-mel2,CH-4-mel3}, the symmetry algebra of the $\kappa$-Minkowski space-time has been realised using an undeformed $\kappa$-Poincar\'e algebra, which has a standard Poincar\'e algebra and deformed co-algebra sectors. The explicit form of the Lorentz generators in the undeformed $\kappa$-Poincar\' e algebra is given by \cite{CH-4-hopf}
\begin{equation}\label{Chap-4-CFK8}
\begin{split}
 M_{ij}=& x_i\partial_j-x_j\partial_i ~,\\
 M_{i0}=& x_i\partial_0\varphi\frac{e^{2A}-1}{2A}-x_0\partial_i\frac{1}{\varphi}+iax_i\partial_k^2\frac{1}{2\varphi}-iax_k\partial_k\partial_i\frac{\gamma}{\varphi},
\end{split} 
\end{equation}
The standard Poincar\'e algebra structure is retained with the help of a new derivative known as Dirac derivative $D_{\mu}$ \cite{CH-4-hopf}. This $D_{\mu}$ transforms as a $4$-vector under the undeformed $\kappa$-Poincar\'e algebra, unlike the standard derivative $\partial_{\mu}$. The components of this $D_{\mu}$ are given as \cite{CH-4-hopf}
\begin{equation}\label{Chap-4-CFK10}
\begin{split}
  D_i=\partial_i\frac{e^{-A}}{\varphi},~~D_0=\partial_0\frac{\sinh A}{A}+ia\partial_k^2\frac{e^{-A}}{2\varphi^2}.
\end{split}
\end{equation}
Using the generators obtained above, the undeformed $\kappa$-Poincar\'e algebra is defined as \cite{CH-4-hopf}
\begin{equation}\label{Chap-4-CFK9}
\begin{split}  
  [M_{\mu\nu},M_{\lambda\rho}]= &M_{\mu\rho}\eta_{\nu\lambda}-M_{\nu\rho}\eta_{\mu\lambda}-M_{\mu\lambda}\eta_{\nu\rho}+M_{\nu\lambda}\eta_{\mu\rho},\\
 [M_{\mu\nu},D_{\lambda}]=&D_{\mu}\eta_{\nu\lambda}-D_{\nu}\eta_{\mu\lambda},~~
 [D_{\mu},D_{\nu}]=0.
\end{split}
\end{equation}
The quadratic Casimir corresponding to this undeformed $\kappa$-Poincar\'e algebra is given as \cite{CH-4-hopf}
\begin{equation}\label{Chap-4-CFK11}
 D_{\mu}D^{\mu}=\partial_k^2\frac{e^{-A}}{2\varphi^2}-\partial_0^2\frac{\sinh^2{A/2}}{(A/2)^2} + \frac{a^2}{4}\left( \partial_k^2\frac{e^{-A}}{2\varphi^2}-\partial_0^2\frac{\sinh^2{A/2}}{(A/2)^2} \right)^2.
\end{equation}
From Eq.(\ref{Chap-4-CFK9}), it is evident that $[M_{\mu\nu},D_{\lambda}D^{\lambda}]=0$ and the above defined Casimir operator remains invariant under the undeformed $\kappa$-Poincar\'e algebra. Therefore, we use this quadratic Casimir operator to construct the invariant Lagrangian for the scalar field in $\kappa$-Minkowski space-time as \cite{CH-4-skp,vishnu1}, i.e., 
\be
\mathcal{L} = \frac{1}{2}\phi(x)D_{\mu}D^{\mu}\phi(x)\label{lag}.
\ee
In the next section, using this deformed Lagrangian, we construct the partition function for studying the finite temperature Casimir effect in $\kappa$-Minkowski space-time. 

\section{Deformed finite temperature Casimir effect}

We calculate the finite temperature corrections to the Casimir energy for a massless real scalar field confined between parallel plates in the $\kappa$-Minkowski space-time. Our analysis commences with constructing the $\kappa$-deformed partition function using the deformed quadratic Casimir operator, such that the Lagrangian remains invariant under the undeformed $\kappa$-Poincar\' e algebra \cite{CH-4-hopf,CH-4-mel2,CH-4-mel3}. Here we employ the Matsubara formalism for bosonic modes to study the finite-temperature Casimir effects \cite{matsubara1,matsubara2}. After computing the partition function, we explicitly obtain the expressions for the free energy and the Casimir force. This approach provides a consistent framework to analyze the combined effects of spatial confinement and thermal fluctuations in the $\kappa$-deformed space-time.

Thus, the relevant partition function, in the Euclidean path integral formalism, is constructed using Eq.(\ref{Chap-4-CFK11}) and $\varphi=e^{-\frac{A}{2}}$ in Eq.(\ref{lag}) as,
\begin{equation}\label{a1}
 Z \equiv \int \mathcal{D} \phi ~ e^{- S_E[\phi]}=\int \mathcal{D} \phi ~ e^{- \frac{1}{2}\int d\tau \int d^3x~\phi(x)\Big(-\partial_i^2 - \partial_{\tau}^2 - \frac{a^2}{3}\partial_{\tau}^4 - \frac{a^2}{4} \partial_i^4 - \frac{a^2}{2}\partial_i^2 \partial_{\tau}^2 \Big) \phi(x)}
\end{equation}
For this particular realisation $\varphi=e^{-\frac{A}{2}}$, the first NC correction term in the $\kappa$-deformed Lagrangian (i.e., from the $\kappa$-deformed quadratic Casimir operator) appears in $a^2$ dependent term. So in this study, we do the calculations valid up to $a^2$ terms.

To incorporate the finite temperature effects, we use the Matsubara formalism, wherein the Euclidean time coordinate of the scalar field satisfies periodic boundary conditions, i.e., $\phi(\tau,x)=\phi(\tau+\beta,x)$. This periodicity is inversely related to the temperature $T$ as $\beta=1/T$, resulting in a discrete set of Matsubara frequency modes given by $\omega_n=\frac{2n\pi}{\beta}$, where $n\in Z$ \cite{matsubara1,matsubara2}. Expanding this field using the Fourier series, we get
\begin{equation}\label{a3}
 \phi(\tau,x)=\sum_{n=-\infty}^{\infty}  e^{i(\omega_n\tau + k\cdot x)}
\end{equation}
Upon evaluating Eq.(\ref{a1}), we get $Z=(\textnormal{det}[\mathcal{A}])^{-1/2}$, where $\mathcal{A}=-\partial_i^2 - \partial_{\tau}^2 - \frac{a^2}{3}\partial_{\tau}^4 - \frac{a^2}{4} \partial_i^4 - \frac{a^2}{2}\partial_i^2 \partial_{\tau}^2$. Now we calculate this det$[\mathcal{A}]$ from its eigenvalue equation $\mathcal{A}\phi=\lambda\phi$. After substituting Eq.(\ref{a3}) in this eigenvalue equation, we get the eigenvalue $\lambda_n$ corresponding to the mode $\omega_n$ as $\lambda_n = k_i^2 + \omega_n^2 - \frac{a^2}{3}\omega_n^4 - \frac{a^2}{4} k_i^4 - \frac{a^2}{2} k_i^2 \omega_n^2$. By taking the product of these eigenvalues, det$[\mathcal{A}]$ is obtained as
\begin{equation}\label{a4}
 \textnormal{det}[\mathcal{A}]=\prod_{n,k} \Big(k_i^2 + \omega_n^2 - \frac{a^2}{3}\omega_n^4 - \frac{a^2}{4} k_i^4 - \frac{a^2}{2} k_i^2 \omega_n^2 \Big).
\end{equation}
Thus, we obtain the expression for the free energy by using Eq.(\ref{a3}) and $Z=(\textnormal{det}[\mathcal{A}])^{-1/2}$ in the definition $ F=-\beta \ln Z$, as 
\begin{equation}\label{a4}
 F = \frac{1}{2\beta}\sum_{n} \int \frac{d^3k}{(2\pi)^3} V \ln \Big(k^2 + \omega_n^2 - \frac{a^2}{3}\omega_n^4 - \frac{a^2}{4} k^4 - \frac{a^2}{2} k^2 \omega_n^2 \Big)
\end{equation}
In the above expression, the spatial momentum coordinates are continuous, so we replace the summation in $k$ using integrals with $V$ being the spatial volume. On re-arranging Eq.(\ref{a4}), the commutative as well as the non-commutative contributions can be separately written as
\begin{equation}\label{a5}
 F = \frac{1}{2\beta}\sum_{n} \int \frac{d^3k}{(2\pi)^3} V \bigg\{\ln \Big(k^2 + \omega_n^2 \Big) + \ln \Big( 1 - \frac{\frac{a^2}{3}\omega_n^4 + \frac{a^2}{4} k^4 + \frac{a^2}{2} k^2\omega_n^2}{k^2 + \omega_n^2}\Big) \bigg\}.
\end{equation}
As the $\kappa$-deformed correction terms are small compared to their commutative counterparts, we can evaluate the second term of Eq.(\ref{a5}) by expanding the logarithmic term up to $a^2$ dependent term  
\begin{equation}\label{a6}
 F = \frac{1}{2\beta} \int \frac{d^3k}{(2\pi)^3} V ~\sum_{n=-\infty}^{\infty} ~\ln \Big(k^2 + \omega_n^2 \Big) -\frac{a^2}{24\beta}\int \frac{d^3k}{(2\pi)^3} V \sum_{n=-\infty}^{\infty}  \Big(\frac{ 4\omega_n^4 + 3k^4 + 6k^2\omega_n^2}{k^2 + \omega_n^2}\Big) 
\end{equation}
In general, the free energy $F$ in Eq.(\ref{a6}) can be decomposed as $F= E_0 + F_T$, where $E_0$ is the zero temperature part and $F_T$ is the temperature dependent part of the deformed free energy. Now evaluating the first term of Eq.(\ref{a6}) using $\displaystyle\sum_{n=-\infty}^{\infty}\frac{1}{n^2+\alpha^2}=\frac{\pi}{\alpha}\coth{\pi\alpha}$, we get the temperature dependent part of the commutative free energy as 
\begin{equation}\label{a7}
 F_T^{(0)} = \frac{V}{\beta} \int \frac{d^3k}{(2\pi)^3} \ln \big(1-e^{-\beta k}\big)
\end{equation}
The second term of Eq.(\ref{a6}) is calculated by converting the summation into contour integrals, using the Mastubara frequency summation for the bosonic modes, i.e., $\frac{1}{\beta}\displaystyle\sum_{n=-\infty}^{\infty} f(i\omega_n) = \oint_C \frac{dz}{2\pi i} f(z) \frac{1}{e^{\beta z}-1}$ \cite{matsubara1,matsubara2} and evaluating this contour integral (as in appendix A), we get the $\kappa$-dependent correction term of the thermal free energy as
\begin{equation}\label{a9}
 F^{(a)} =-\frac{a^2V}{24} \int \frac{d^3k}{(2\pi)^3} \frac{k^3}{2}\coth{\left(\frac{k\beta}{2}\right)}
\end{equation}
Now, consider two parallel plates fixed at $z=0$ and $z=L$ and by imposing the Dirichlet boundary condition, we find the momentum component along the $z$ direction to be $k_m=\frac{m\pi}{L}$ where $m=1,2,... $. Thus, we get $k=\sqrt{k_{\perp}^2+k_m^2}$ and $dk_m$ becomes $\frac{dk_m}{2\pi}\to\frac{1}{L}\displaystyle\sum_{m=1}^{\infty}$. By rewriting the volume $V$ in terms of $L$ and area $A$, we get the Casimir free energy per unit area, i.e., $\mathcal{F}=F/A$, as
\begin{equation}\label{a10}
 \mathcal{F}_T = \frac{1}{2\beta} \sum_{m=1}^{\infty}\int \frac{d^2k_{\perp}}{(2\pi)^2} \ln \Big(1-e^{-\beta\sqrt{k_{\perp}^2 + \frac{m^2\pi^2}{L^2}}}  \Big) - \sum_{m=1}^{\infty}\frac{a^2}{48}\int\frac{d^2k_{\perp}}{(2\pi)^2}  \left(k_{\perp}^2+\frac{m^2\pi^2}{L^2}\right)^{3/2}\coth{\left(\frac{\beta\sqrt{k_{\perp}^2+\frac{m^2\pi^2}{L^2}}}{2}\right)}
 \end{equation}
Note that the explicit form of $k$ changes as the imposed boundary condition changes (as shown in \cite{oikonomou1}). Similarly, if we consider a compact dimension (say along the $y$ direction), then we obtain $k=\sqrt{k_{x}^2 + \left(\frac{2\tilde{m}\pi}{R}\right)^2 + \frac{m^2\pi^2}{L^2}}$, where $\tilde{m}$ is an integer and $R$ is the length of the compact dimension. All these modifications seen in $k$ due to changes in the boundary conditions or due to the presence of compact dimensions, lead to the modification of $\mathcal{F}_T$, given in Eq.(\ref{a10}). Hence, changes in the boundary conditions or the presence of compact dimensions affect the final expression for the thermal Casimir energy, even under $\kappa$-deformation. Such modifications can also be observed if we consider non-trivial field configurations which induce anti-periodic boundary conditions on the bosonic field, as in \cite{ok3}.
 
\begin{figure}[!htb]\centering
    \includegraphics[width=0.65\linewidth]{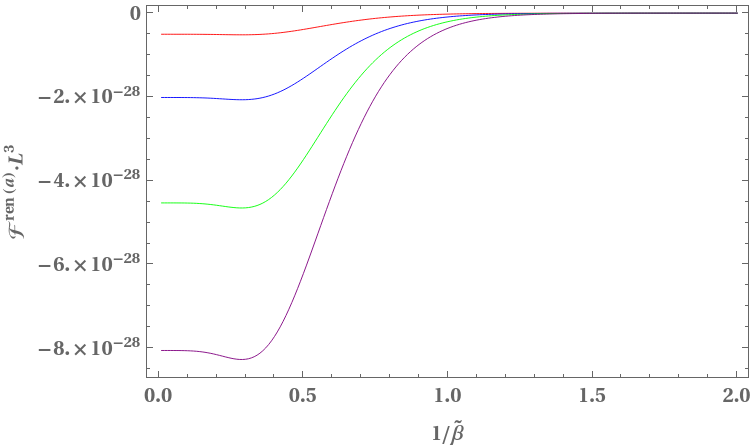}
        \caption{Variation of $\mathcal{F}^{ren~(a)}$ against $1/\tilde{\beta}$ for certain fixed values of $a/L$: (i) $a/L=0.25\times 10^{-12}$(red), $a/L=0.5\times 10^{-12}$(blue), $a/L=0.75\times 10^{-12}$(green), $a/L=1\times 10^{-12}$(purple). Here we obtain this plot by evaluating the summation in Eq.(\ref{a18}) with a truncation of $m=500$}
    \label{fig:plot2}
\end{figure}

First, we evaluate the commutative term of Eq.(\ref{a10}) by expanding the logarithmic term using $\ln{(1-x)}=-\displaystyle\sum_{n=1}^{\infty}\frac{x^n}{n}$ and we get $\mathcal{F}_{T}^{~(0)} =  -\frac{1}{2\pi\beta^3} \displaystyle\sum_{m=1}^{\infty} \displaystyle\sum_{n=1}^{\infty} \frac{1}{n} \displaystyle\int_{m\pi/L}^{\infty} d\bar{k}~\bar{k} e^{-n\beta \bar{k}}$, where $\bar{k}=\sqrt{k_{\perp}^2+m^2\pi^2/L^2}$. Carrying out this integral and evaluating the summation over $n$, we arrive at the standard expression for $\mathcal{F}_T^{(0)}$ as \cite{zhang1}
\begin{equation}\label{a12}
 \mathcal{F}_{T}^{~(0)} =  -\frac{\pi^2}{32L^3} \sum_{m=1}^{\infty} \left( \frac{\coth{m\tilde{\beta}}}{m^3\tilde{\beta}^3} + \frac{1}{m^2\tilde{\beta}^2 \sinh^2{m\tilde{\beta}}}\right) + \frac{\zeta(3)}{4\pi\beta^3}
\end{equation}
where $\tilde{\beta}=\pi\beta/2L$. By substituting $\bar{k}=\sqrt{k_{\perp}^2+m^2\pi^2/L^2}$ and using the identity $\coth{x}=1+\tfrac{2}{e^{2x}-1}$ in the second term of Eq.(\ref{a10}), the temperature-dependent NC correction part of the Casimir free energy becomes
\begin{equation}\label{a13}
    \mathcal{F}_T^{(a)}= - \frac{a^2}{48} \sum_{m=1}^{\infty} \int_{m\pi/L}^{\infty} d\bar{k}~\frac{\bar{k}^4}{e^{\beta \bar{k}}-1}
\end{equation}
After evaluating this integral (as shown in Appendix B), we obtain the NC parts of thermal Casimir free energy in $\kappa$-Minkowski space-time as
\begin{equation}\label{a15}
\begin{split}
    \mathcal{F}_T^{(a)} = -\frac{a^2\pi^4}{96L^5} \sum_{m=1}^{\infty} \bigg(& \frac{3\coth{m\tilde{\beta}}}{4m^5\tilde{\beta}^5} + \frac{3}{4m^4\tilde{\beta}^4 \sinh^2{m\tilde{\beta}}} + \frac{3\coth{m\tilde{\beta}}}{4m^3\tilde{\beta}^3 \sinh^2{m\tilde{\beta}}} + \frac{\coth^2{m\tilde{\beta}}}{2m^2\tilde{\beta}^2 \sinh^2{m\tilde{\beta}}} \\
    &+ \frac{1}{4m^4\tilde{\beta}^4 \sinh^4{m\tilde{\beta}}} + \frac{\coth^2{m\tilde{\beta}}}{4m\tilde{\beta} \sinh^2{m\tilde{\beta}}}  + \frac{\coth{m\tilde{\beta}}}{2m\tilde{\beta} \sinh^4{m\tilde{\beta}}}\bigg) + \frac{a^2\zeta(5)}{4\pi\beta^5}
\end{split}
\end{equation}
Therefore, by combining Eq.(\ref{a12}) and Eq.(\ref{a15}), we get the temperature dependent part of the Casimir free energy in $\kappa$-Minkowski space-time, valid up to $a^2$ term, as
\begin{equation}\label{a16}
\begin{split}
    \mathcal{F}_T = \; &
    -\frac{\pi^2}{32 L^3} \sum_{m=1}^\infty \left(
        \frac{\coth m\tilde{\beta}}{m^3 \tilde{\beta}^3} 
        + \frac{1}{m^2 \tilde{\beta}^2 \sinh^2 m\tilde{\beta}}
    \right) 
    + \frac{\zeta(3)}{4 \pi \beta^3} + \frac{a^2 \zeta(5)}{4 \pi \beta^5}  \\
    &
    - \frac{a^2 \pi^4}{96 L^5} \sum_{m=1}^\infty \Bigg(
        \frac{3 \coth m\tilde{\beta}}{4 m^5 \tilde{\beta}^5} 
        + \frac{3}{4 m^4 \tilde{\beta}^4 \sinh^2 m\tilde{\beta}} 
        + \frac{3 \coth m\tilde{\beta}}{4 m^3 \tilde{\beta}^3 \sinh^2 m\tilde{\beta}} \\
    & 
        + \frac{\coth^2 m\tilde{\beta}}{2 m^2 \tilde{\beta}^2 \sinh^2 m\tilde{\beta}} 
        + \frac{1}{4 m^4 \tilde{\beta}^4 \sinh^4 m\tilde{\beta}} 
        + \frac{\coth^2 m\tilde{\beta}}{4 m \tilde{\beta} \sinh^2 m\tilde{\beta}} 
        + \frac{\coth m\tilde{\beta}}{2 m \tilde{\beta} \sinh^4 m\tilde{\beta}}
    \Bigg) \\
\end{split}
\end{equation}
From Eq.(\ref{a16}), we observe that $\mathcal{F}_T$ depends only on the plate separation and the temperature, apart from the $\kappa$-deformation parameter $a$. So to renormalise this thermal Casimir free energy, we need to analyse the asymptotic behaviour of $\mathcal{F}_T$ for infinite plate separation distance (i.e., $L\to\infty$). Thus by series expanding $\mathcal{F}_T$ under this conditions, we obtain $ -\frac{\pi^2L}{90\beta^4} + \frac{\zeta(3)}{4\pi\beta^3} - \frac{a^2\pi^4L}{378\beta^6} + \frac{a^2\zeta(5)}{4\pi\beta^5}$. Next, subtracting this asymptotic limit from Eq.(\ref{a16}) (as in \cite{zhang1}), we obtain the renormalised thermal Casimir free energy as
\begin{equation}\label{a18}
\begin{split}
    \mathcal{F}^{ren}_T = \; & 
    -\frac{\pi^2}{32L^3} \sum_{m=1}^{\infty} \left( 
        \frac{\coth m\tilde{\beta}}{m^3 \tilde{\beta}^3} 
        + \frac{1}{m^2 \tilde{\beta}^2 \sinh^2 m\tilde{\beta}}
    \right) 
    + \frac{\pi^2 L}{90 \beta^4} + \frac{a^2 \pi^4 L}{378 \beta^6}  \\
    & 
    - \frac{a^2 \pi^4}{96 L^5} \sum_{m=1}^{\infty} \bigg(
        \frac{3 \coth m\tilde{\beta}}{4 m^5 \tilde{\beta}^5} 
        + \frac{3}{4 m^4 \tilde{\beta}^4 \sinh^2 m\tilde{\beta}} 
        + \frac{3 \coth m\tilde{\beta}}{4 m^3 \tilde{\beta}^3 \sinh^2 m\tilde{\beta}} \\
    & 
        + \frac{\coth^2 m\tilde{\beta}}{2 m^2 \tilde{\beta}^2 \sinh^2 m\tilde{\beta}}
        + \frac{1}{4 m^4 \tilde{\beta}^4 \sinh^4 m\tilde{\beta}} 
        + \frac{\coth^2 m\tilde{\beta}}{4 m \tilde{\beta} \sinh^2 m\tilde{\beta}} 
        + \frac{\coth m\tilde{\beta}}{2 m \tilde{\beta} \sinh^4 m\tilde{\beta}}
    \bigg) \\
\end{split}
\end{equation}
For a complete understanding of the thermal Casimir effect in $\kappa$-Minkowski space-time, we also need to incorporate the zero-temperature contributions into it. From Eq.(\ref{a9}), we obtain the renormalised zero-temperature Casimir free energy as $\varepsilon_0^{ren}=-\frac{\pi^2}{1440L^3}-\frac{a^2\pi^4}{120960L^5}$. Hence the total renormalised Casimir free energy in $\kappa$-Minkowski space-time is $\mathcal{F}^{ren}=\varepsilon_0^{ren}+\mathcal{F}_T^{ren}$, i.e.,
\begin{equation}\label{b3}
\begin{split}
    \mathcal{F}^{ren} = \; & 
    -\frac{\pi^2}{1440L^3}-\frac{\pi^2}{32L^3} \sum_{m=1}^{\infty} \left( 
        \frac{\coth m\tilde{\beta}}{m^3 \tilde{\beta}^3} 
        + \frac{1}{m^2 \tilde{\beta}^2 \sinh^2 m\tilde{\beta}}
    \right) + \frac{\pi^2 L}{90 \beta^4} \\
    &
    -\frac{a^2\pi^4}{120960L^5}  - \frac{a^2 \pi^4}{96 L^5} \sum_{m=1}^{\infty} \bigg(
        \frac{3 \coth m\tilde{\beta}}{4 m^5 \tilde{\beta}^5} 
        + \frac{3}{4 m^4 \tilde{\beta}^4 \sinh^2 m\tilde{\beta}} 
        + \frac{3 \coth m\tilde{\beta}}{4 m^3 \tilde{\beta}^3 \sinh^2 m\tilde{\beta}} \\
    & 
        + \frac{\coth^2 m\tilde{\beta}}{2 m^2 \tilde{\beta}^2 \sinh^2 m\tilde{\beta}}
        + \frac{1}{4 m^4 \tilde{\beta}^4 \sinh^4 m\tilde{\beta}} 
        + \frac{\coth^2 m\tilde{\beta}}{4 m \tilde{\beta} \sinh^2 m\tilde{\beta}} 
        + \frac{\coth m\tilde{\beta}}{2 m \tilde{\beta} \sinh^4 m\tilde{\beta}}
    \bigg) + \frac{a^2 \pi^4 L}{378 \beta^6} \\
\end{split}
\end{equation}
From Fig.(\ref{fig:plot2}), it is observed that the NC part of the Casimir free energy, $\mathcal{F}^{ren~(a)}$, remains negative, just like its commutative counterpart and this $\mathcal{F}^{ren~(a)}$ is found to become more negative for larger $a/L$ values. Hence, the total Casimir free energy remains negative in the $\kappa$-deformed space-time. We also notice that this correction term becomes significant in the low temperature regions. 

In the low-temperature regime we have $\tilde{\beta}>>1$, (i.e., $T<<\pi/2L$). Using these conditions in Eq.(\ref{a18}), the low temperature limit for the temperature dependent part of the Casimir free energy in $\kappa$-deformed space-time is obtained as 
\begin{equation}\label{b1}
\begin{split}
     \mathcal{F}^{low}_T \simeq & -\frac{\zeta(3)}{4\pi\beta^3} + \frac{\pi^2 L}{90\beta^4} - \frac{e^{-\pi\beta/L}}{16L\beta^2} - \frac{a^2\zeta(5)}{4\pi\beta^5} + \frac{a^2\pi^4L}{378\beta^6} - a^2\left(\frac{\pi^2}{6L^3\beta^2} + \frac{\pi^4}{12L^4\beta} \right)e^{-2\pi\beta/L}\\
     &- a^2\left(\frac{1}{2L\beta^4} + \frac{\pi}{4L^2\beta^3} + \frac{\pi^2}{12L^3\beta^2} + \frac{\pi^3}{48L^4\beta} \right)e^{-\pi\beta/L}
\end{split}
\end{equation}
The NC part of the thermal Casimir free energy, i.e., $\mathcal{F}^{ren~(a)}_T$, remains negative throughout in the low temperature limit and thus it is in agreement with that seen in Fig.(\ref{fig:plot2}). At the zero temperature, $\mathcal{F}^{ren}_T$ also vanishes to zero, from Eq.(\ref{b1}). Thus, at zero temperature, the total Casimir free energy will be contributed by $\varepsilon^{ren}_0$ only. The total Casimir pressure associated with the renormalised Casimir free energy can be calculated from $\mathcal{P}=-\frac{\partial\mathcal{F}^{ren}}{\partial L}$. Substituting Eq.(\ref{b3}) in this, we get the total Casimir pressure in $\kappa$-deformed space-time as
\begin{equation}\label{a19}
\begin{split}
     \mathcal{P} = & -\frac{\pi^2}{480L^4} + \frac{\pi^2}{16L^4} \sum_{m=1}^{\infty} \frac{\coth{m\tilde{\beta}}}{m\tilde{\beta}\sinh^2{m\tilde{\beta}}} - \frac{\pi^2}{90\beta^4}\\
     &
     -\frac{a^2\pi^4}{24192L^6}+\frac{a^2\pi^4}{1536L^6} \sum_{m=1}^{\infty} \left(\frac{33 + 26\cosh{2m\tilde{\beta}} + \cosh{4m\tilde{\beta}}}{\sinh^6{m\tilde{\beta}}}\right) - \frac{a^2\pi^4}{378\beta^6}
\end{split}
\end{equation}

\begin{figure}[!htb]
    \centering
    \includegraphics[width=0.65\linewidth]{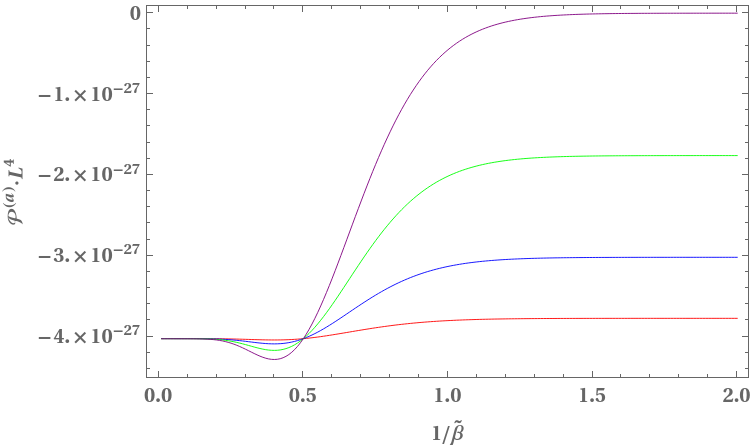}
        \caption{Plots show the variation of $\mathcal{P}^{(a)}$ against $1/\tilde{\beta}$ for certain fixed values of $a/L$: (i) $a/L=0.25\times 10^{-12}$(red), $a/L=0.5\times 10^{-12}$(blue), $a/L=0.75\times 10^{-12}$(green), $a/L=1\times 10^{-12}$(purple). To obtain this, we evaluating the summation in Eq.(\ref{a19}) with a truncation of $m=500$}
    \label{fig:plot4}
\end{figure}

In Fig.(\ref{fig:plot4}), $\mathcal{P}^{(a)}$ initially becomes more negative for larger $a/L$ values in the low temperature region and as the temperature rises the thermal correction terms dominate over the zero-temperature correction terms so that $\mathcal{P}^{(a)}$ grows less negative for larger $a/L$ values. However, the plot shows that $\mathcal{P}^{(a)}$ is totally negative, leading to an attractive nature of the NC part of the Casimir pressure. Thus, it is evident that the total Casimir pressure in the $\kappa$-deformed space-time is also attractive. In addition, we obtain a bound on the deformation parameter $a$ from the expression for the zero-temperature Casimir pressure in $\kappa$-deformed space-time. Comparing $\mathcal{P}_{T=0}=-\frac{\pi^2}{480L^4}-\frac{a^2\pi^4}{24192L^6}$ with the result reported in \cite{bound}, having Casimir force $(1.22\pm 0.18)\times 10^{-27}~N/m^2$ and plate separation $1\mu m$, we get an upper bound as $a<10^{-18}m$. Hence, we expect the $\kappa$-deformed corrections to the Casimir effect to become significant when $a/L\leq10^{-12}$.  

\begin{figure}[!htb]
    \centering
    \includegraphics[width=0.65\linewidth]{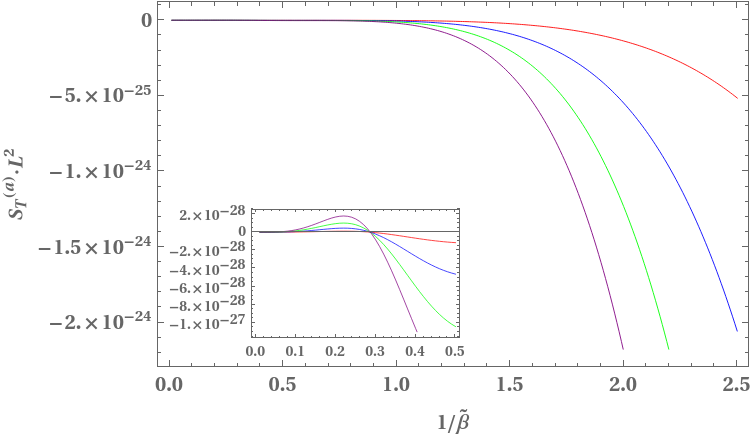}
        \caption{Plots show the variation of $\mathcal{S}^{(a)}_T$ against $1/\tilde{\beta}$ for certain fixed values of $a/L$: (i) $a/L=0.25\times 10^{-12}$(red), $a/L=0.5\times 10^{-12}$(blue), $a/L=0.75\times 10^{-12}$(green), $a/L=1\times 10^{-12}$(purple). This plot is obtained by evaluating the summation in Eq.(\ref{a18}) with a truncation of $m=500$}
    \label{fig:plot5}
\end{figure}

The entropy associated with the thermal Casimir free energy is obtained from Eq.(\ref{a18}) using $\mathcal{S}_T=-\frac{\partial \mathcal{F}^{ren}_T}{\partial T}$, as
\begin{equation}\label{a21}
\begin{split}
     \mathcal{S}_T = & \frac{\pi^2\beta}{32L^3} \sum_{m=1}^{\infty} \left( \frac{3\coth{m\tilde{\beta}}}{m^3\tilde{\beta}^3} + \frac{3}{m^2\tilde{\beta}^2 \sinh^2{m\tilde{\beta}}} +  \frac{2\coth{m\tilde{\beta}}}{m\tilde{\beta} \sinh^2{m\tilde{\beta}}}\right) - \frac{2\pi^2 L}{45\beta^3}
     +\frac{a^2\pi^4\beta}{96L^5} \sum_{m=1}^{\infty} \bigg(\frac{15\coth{m\tilde{\beta}}}{4m^5\tilde{\beta}^5} \\
     &+ \frac{15}{4m^4\tilde{\beta}^4 \sinh^2{m\tilde{\beta}}}
     + \frac{15\coth{m\tilde{\beta}}}{4m^3\tilde{\beta}^3 \sinh^2{m\tilde{\beta}}} + \frac{5\coth{m\tilde{\beta}}}{2m^2\tilde{\beta}^2 \sinh^2{m\tilde{\beta}}}
    + \frac{5\coth^3{m\tilde{\beta}}}{m\tilde{\beta} \sinh^2{m\tilde{\beta}}} + \frac{\coth^4{m\tilde{\beta}}}{2 \sinh^2{m\tilde{\beta}}}  \\
    &+ \frac{5}{4m^2\tilde{\beta}^2 \sinh^4{m\tilde{\beta}}} + \frac{5\coth{m\tilde{\beta}}}{2m\tilde{\beta} \sinh^4{m\tilde{\beta}}} + \frac{11\coth^2{m\tilde{\beta}}}{4 \sinh^4{m\tilde{\beta}}} + \frac{1}{2\sinh^6{m\tilde{\beta}}} \bigg) - \frac{a^2\pi^4 L}{63\beta^5}
\end{split}
\end{equation}
Fig.(\ref{fig:plot5}) shows that while the $\kappa$-deformed correction to the entropy, i.e., $\mathcal{S}^{(a)}_T$ is initially positive and vanishes at a characteristic temperature $T_{min}$ where the summation terms exactly compensate for the $\frac{a^2\pi^4 L}{63\beta^5}$ contribution. In particular, for $T > T_{min}$, the term $\frac{a^2\pi^4 L}{63\beta^5}$ dominates, leading to negative entropy values and a violation of the second law of thermodynamics. This apparent local violation of the second law of thermodynamics suggests the presence of some non-equilibrium effects associated with the $\kappa$-deformed correction part of the Casimir effect. However, the total entropy of this Casimir system in the $\kappa$-deformed space-time, i.e., $\mathcal{S}_{T}$, is positive and consequently this is thermodynamically stable.

\begin{figure}[!htb]
    \centering
    \includegraphics[width=0.65\linewidth]{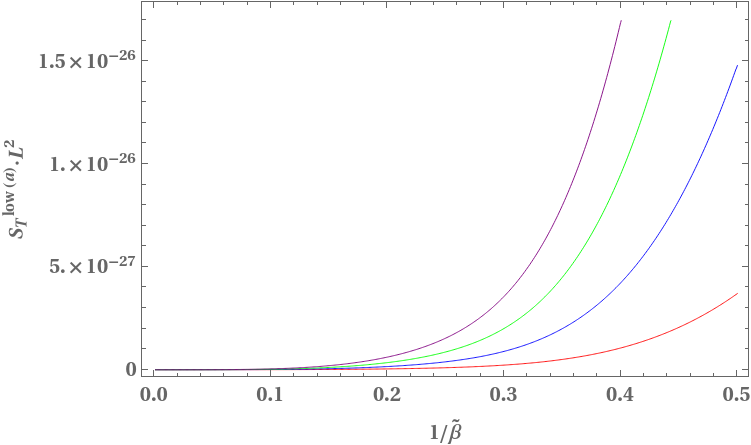}
        \caption{Variation of $\mathcal{S}^{low~(a)}_T$ against $1/\tilde{\beta}$ in the low temperature limit for certain $a/L$ values:  (i) $a/L=0.25\times 10^{-12}$(red), $a/L=0.5\times 10^{-12}$(blue), $a/L=0.75\times 10^{-12}$(green), $a/L=1\times 10^{-12}$(purple)}
        \label{fig:plot6}
\end{figure}
From Eq.(\ref{a21}), we find the asymptotic behaviour of the $\kappa$-deformed entropy associated with the Casimir effect in the low temperature limit as
\begin{equation}\label{b2}
\begin{split}
     \mathcal{S}^{low} \simeq & ~\frac{3\zeta(3)}{4\pi\beta^2} - \frac{2\pi^2 L}{45\beta^3} + \left(\frac{3}{8L\beta} + \frac{\pi}{2L^2}\right)e^{-\pi\beta/L} + \frac{5a^2\zeta(5)}{4\pi\beta^4} + \frac{a^2\pi^4L}{63\beta^5} + \frac{a^2\pi\beta}{3L^5}e^{-3\pi\beta/L} +\\
     & a^2\left(\frac{5\pi^2}{12L^3\beta} + \frac{5\pi^3}{8L^4} + \frac{44\beta}{L^5} \right)e^{-2\pi\beta/L}
      + a^2\left(\frac{15}{6L\beta^3} + \frac{5\pi}{4L^2\beta^2} + \frac{5\pi^2}{12L^3\beta} + \frac{5\pi^3}{12L^4} + \frac{2\beta}{L^5}\right)e^{-\pi\beta/L}
\end{split}
\end{equation}
Both Eq.(\ref{a21}) and Eq.(\ref{b2}) reduce to zero when the temperature vanishes (i.e., at $T=0$). Thus, we find that the entropy expression corresponding to the $\kappa$-deformed thermal Casimir free energy is in accordance with the Nernst heat theorem. Further from Fig.(\ref{fig:plot6}), we observe that the $\kappa$-deformed part of the entropy increases as the temperature rises slowly, showing the low temperature part is in total agreement with the second law of thermodynamics. Further, using the definition $\mathcal{U}=-T^2\frac{\partial}{\partial T}\big(\frac{\mathcal{F}^{ren}_T}{T}\big)$ and Eq.(\ref{a18}), we find the internal energy associated with the thermal Casimir free energy in $\kappa$-deformed space-time as
\begin{equation}\label{a20}
\begin{split}
\mathcal{U} =\;& \frac{\pi^2}{16L^3} \sum_{m=1}^{\infty} \left( 
\frac{\coth{(m\tilde{\beta})}}{m^3\tilde{\beta}^3} 
+ \frac{1 + m\tilde{\beta}\coth{(m\tilde{\beta})}}{m^2\tilde{\beta}^2 \sinh^2{(m\tilde{\beta})}} 
\right) 
- \frac{\pi^2 L}{30\beta^4} + \frac{a^2\pi^4 L}{378\beta^6} \\
& + \frac{a^2\pi^4}{96L^5} \sum_{m=1}^{\infty} \bigg(
\frac{3\coth{(m\tilde{\beta})}}{4m^5\tilde{\beta}^5} 
+ \frac{3}{4m^4\tilde{\beta}^4 \sinh^2{(m\tilde{\beta})}} 
+ \frac{3\coth{(m\tilde{\beta})}}{4m^3\tilde{\beta}^3 \sinh^2{(m\tilde{\beta})}} \\
& + \frac{\coth^2{(m\tilde{\beta})}}{2m^2\tilde{\beta}^2 \sinh^2{(m\tilde{\beta})}} 
+ \frac{1}{4m^4\tilde{\beta}^4 \sinh^4{(m\tilde{\beta})}} 
+ \frac{\coth^2{(m\tilde{\beta})}}{4m\tilde{\beta} \sinh^2{(m\tilde{\beta})}} 
+ \frac{\coth{(m\tilde{\beta})}}{2m\tilde{\beta} \sinh^4{(m\tilde{\beta})}}
\bigg) 
\end{split}
\end{equation}
From Eq.(\ref{a18}), Eq.(\ref{a21}) and Eq.(\ref{a20}) we find that they satisfy the standard thermodynamic relation $\mathcal{U}=\mathcal{F}^{ren}_T-T\mathcal{S}$, even under the $\kappa$-deformation. 

\section{Stefan-Boltzmann law in \texorpdfstring{$\kappa$}{kappa}-Minkowski space-time}

The partition function developed in the previous section has been used to obtain the free energy of black body radiation in the $\kappa$-Minkowski space-time. Using Eq.(\ref{a7}) and Eq.(\ref{a9}), we write down the free energy density of black body radiation, valid up to $a^2$ term, as
\begin{equation}\label{c1}
 \mathcal{F}_{bb} = \frac{1}{\beta} \int \frac{d^3k}{(2\pi)^3} \ln \big(1-e^{-\beta k}\big) -\frac{a^2}{24} \int \frac{d^3k}{(2\pi)^3} \frac{k^3}{2}\coth{\left(\frac{k\beta}{2}\right)}
\end{equation}
Solving the above integrals using $\displaystyle\int_0^{\infty}dk\frac{k^s}{e^{k}-1}=\Gamma(s+1)~\zeta(s+1)$, and after some calculational simplification, we find the $\kappa$-deformed energy density associated with black body radiation as
\begin{equation}\label{c2}
 \mathcal{F}_{bb} = -\frac{\pi^2}{90\beta^4} - \frac{a^2\pi^4}{378\beta^6}
\end{equation}
The above result corresponds to the Stefan-Boltzmann relation in $\kappa$-deformed space-time. Note that the first term in the above equation is the energy density of black body radiation in commutative space-time. The second term represents the first leading-order correction, which comes due to the non-commutativity of space-time.
\begin{figure}[!htb]
    \centering
    \includegraphics[width=0.6\linewidth]{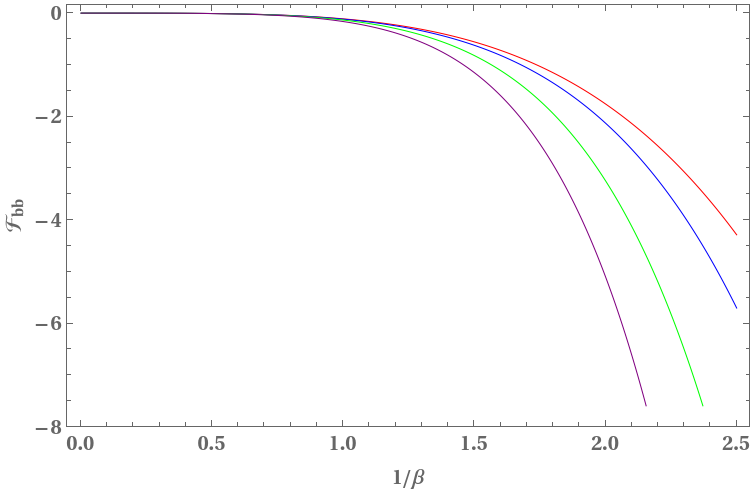}
        \caption{Plot of $\mathcal{F}_{bb}$ against $1/{\beta}$ for certain $a$ values: (i) $a=0$ (red), $a=0.15$ (blue), $a=0.30$ (green), $a=0.45$ (purple)}
    \label{fig:plot7}
\end{figure}
Here we notice that the second-order NC correction term depends on the temperature as $T^6$. One significant consequence of this non-commutativity is the reduction in the energy density of black body radiation as seen in Fig.(\ref{fig:plot7}). This decrease in energy density reflects how non-commutativity imposes a subtle correction to the standard Stefan-Boltzmann law, signaling new physics beyond the classical continuous space-time framework.

Moreover, these terms given in Eq.(\ref{c2}) also appear in the renormalised thermal Casimir free energy expression, i.e., Eq.(\ref{a18}). Such $T^4$ dependent term in the renormalised Casimir free energy expression has been shown to appear in various studies \cite{geyer,zhang1}. Apart from this $T^4$ dependent term, the $\kappa$-deformed non-commutativity induces a $T^6$ dependent term in Eq.(\ref{a18}). From the study and analysis in the previous section, we find that these terms play a crucial role in the thermal behaviour of the Casimir effect.

\section{Conclusion}

We have investigated the finite temperature Casimir effect in the $\kappa$-Minkowski space-time. Using the $\kappa$-deformed Casimir operator, the $\kappa$-deformed scalar field Lagrangian, invariant under the undeformed $\kappa$-Poincaré algebra, has been constructed, and the corresponding partition function has been obtained. By employing the Matsubara formalism and implementing the Dirichlet boundary condition, we have analysed the thermal behavior of the Casimir effect, including Casimir free energy, Casimir force, entropy and internal energy. Our results reveal some interesting modifications to the Casimir effect due to the NC nature of space-time, particularly in the presence of thermal fluctuations, and all these results smoothly reduce to the standard results, in the vanishing limit of $a$. 

The $\kappa$-deformed zero-temperature Casimir energy remains negative and the correction term scales with plate distance as $1/L^5$, which is in agreement with the results of \cite{CH-4-pinto,CH-4-skp}, where the zero-temperature $\kappa$-deformed Casimir energy has been derived separately using the deformed dispersion relation of the $\kappa$-Poincaré algebra and from the Green's function approach corresponding to the $\kappa$-deformed scalar field. Such $1/L^5$ dependent correction term has also been reported while studying the Casimir effect in Moyal space-time \cite{CH-4-casadio}, Snyder space-time \cite{snyder}, in the context of generalised uncertainty principle (GUP) \cite{panella} and from the modified dispersion relation of doubly special relativity (DSR) \cite{davis}. Thus, intuitively, one can expect the appearance of the $1/L^5$ term with a quadratic dependence on the minimal length scale when studying the Casimir effect in the presence of a minimal length scale. So in general, the higher-order minimal length scale dependent correction term would scale as $1/L^n$ signaling the predominance of the space-time non-commutativity at small plate separations.

The $\kappa$-deformed corrections to the thermal Casimir free energy exhibit a non-trivial interplay between the intrinsic NC length scale, geometric scale (plate separation), and the thermal scale of the system. The influence of $\kappa$-deformation becomes increasingly significant in the low-temperature regime, where quantum fluctuations dominate over thermal contributions. The $\kappa$-dependent terms modify the standard Casimir free energy to enhance their relative contribution compared to the standard result. A key observation is that the $\kappa$-deformed corrections to the thermal Casimir free energy remain negative across all temperature regimes and the associated Casimir force retains its attractive nature. The attractive character of the force aligns with the well-known result in commutative space-time, but now incorporates non-perturbative corrections stemming from the $\kappa$-deformation. The persistence of an attractive Casimir force under the $\kappa$-deformation reinforces the robustness of such quantum vacuum effects in the NC space-time structure.

Future investigations could explore whether this deformation-induced behavior leads to observable deviations from standard predictions in high-precision Casimir experiments, particularly in scenarios involving low-temperature and sub-micron separations. Using experimental observations \cite{bound}, we place an improved upper bound on the $\kappa$-deformation parameter as $a\leq 10^{-18}$m, compared to our previous results \cite{CH-4-skp,vishnu}. Interestingly, the $\kappa$-deformation leads to a slight measurable enhancement of the Casimir effect, critically dependent on the ratio $a/L$. This $a/L$ ratio potentially emerges as a crucial experimental parameter in the Casimir effect studies and becomes particularly significant when it approaches experimentally accessible scales (i.e., $a/L\simeq 10^{-12}$), where the interplay between NC geometry and boundary conditions may lead to detectable deviations from standard QFT predictions at finite temperatures. 

The total entropy remains positive, ensuring thermodynamic stability in the $\kappa$-Minkowski space-time. The Nernst heat theorem also remains intact under $\kappa$-deformation as the low-temperature limit of the Casimir entropy vanishes at zero temperature. However, the $\kappa$-deformed corrected part of the Casimir entropy displays an intriguing behavior, including a local violation of the second law of thermodynamics for certain temperature regimes (as seen in Fig.(\ref{fig:plot5})) where the entropy becomes negative. This suggests the possible presence of non-equilibrium effects induced by the NC corrections. 

The study also examined black body radiation in $\kappa$-Minkowski space-time, revealing a modified Stefan-Boltzmann law where the energy density acquires an additional temperature-dependent correction scaling as $T^6$ due to the $\kappa$-deformation of underlying space-time. Similar $T^6$ dependent second order correction term has also appeared in \cite{kappa-black}, for certain orderings compatible with $\kappa$-Poincar\'e algebra. For a different realization, the $a$ dependent correction to the modified Stefan-Boltzmann relation has been shown to have the $T^5$ dependence \cite{suman}. These deviations show how the minimal length scale can alter high-temperature thermodynamics, suggesting observable signatures of NC geometry in thermal radiation phenomena.

However, the calculations become slightly more intricate if we consider a fermionic field, where we need to use the MIT bag model to set boundary conditions, as in \cite{ok4}, to obtain the corresponding eigen frequency for calculating the Casimir energy between parallel plates. Since the fermionic field satisfies the anti-periodic boundary conditions, the resulting Matsubara frequency modes are given by $\omega_n=\frac{(2n+1)\pi}{\beta}$ \cite{matsubara1,matsubara2}. Additionally, we need to consider the correction terms coming from the $\kappa$-deformed Lagrangian for Dirac field \cite{kappa-dirac}. Incorporating these modifications in our study changes the final results of thermal Casimir energy, Casimir force, and entropy. Studies along these directions are in progress and will be reported soon.

\section*{Acknowledgements}

SKP thanks Dr. Mathias Bostr\"om for useful discussions and support. SKP acknowledges the support from the project No. 2022/47/P/ST3/01236 co-funded by the National Science Centre and the European Union's Horizon 2020 research and innovation programme under the Marie Sk{\l}odowska-Curie grant agreement No. 945339. The part of research activity of SKP took place at the ``ENSEMBLE3-Center of Excellence for nanophotonics, advanced materials, and novel crystal growth-based technologies" project (GA No. MAB/2020/14) carried out under the International Research Agenda programs of the Foundation for Polish Science that are co-financed by the European Union under the European Regional  Development Fund and the European Union Horizon 2020 research and innovation program Teaming for Excellence (GA. No. 857543) for supporting this work. Research contributions of SKP to this manuscript were created as part of the project of the Minister of Science and Higher Education ``Support for the activities of Centers of Excellence established in Poland under the Horizon 2020 program" under contract No. MEiN/2023/DIR/3797.

\bigskip

\begin{appendices}

\section{}

\begin{figure}[!htb]
    \centering
    \includegraphics[width=0.5\linewidth]{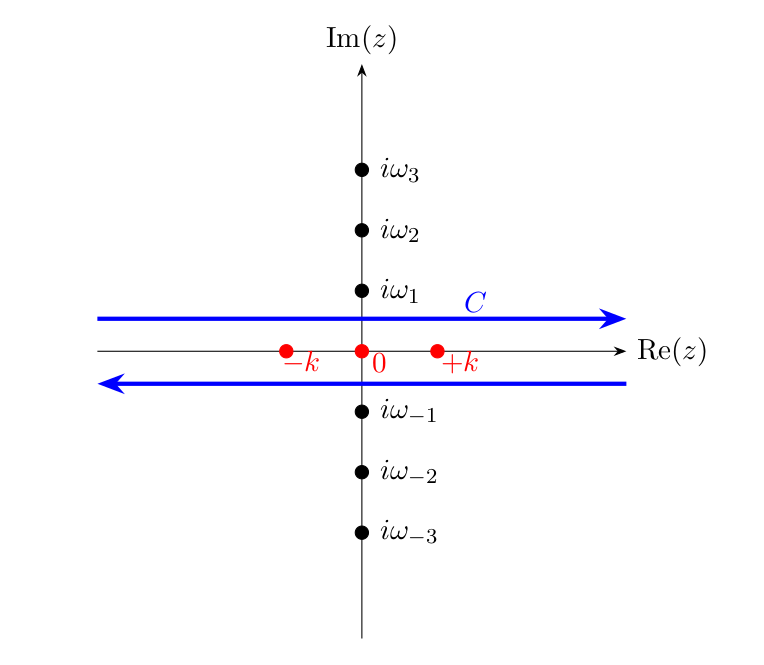}
        \caption{Integration contour $C$ consisting of simple poles at $z=-k,0,+k$}
    \label{fig:plot-a}
\end{figure}

The summation over bosonic Matsubara frequencies can be evaluated using complex contour integration techniques as $\frac{1}{\beta}\displaystyle\sum_{n=-\infty}^{\infty} f(i\omega_n) = \oint_C \frac{dz}{2\pi i} f(z) \frac{1}{e^{\beta z}-1}$.  From the integrand $(e^{\beta z}-1)^{-1}$, we get simple poles located precisely at the Matsubara frequencies as $z = i\frac{2n\pi}{\beta}\equiv i\omega_n$ (where $n=\pm 1,\pm 2$,..), along the imaginary axis. 
\begin{figure}[!htb]
    \centering
    \includegraphics[width=0.5\linewidth]{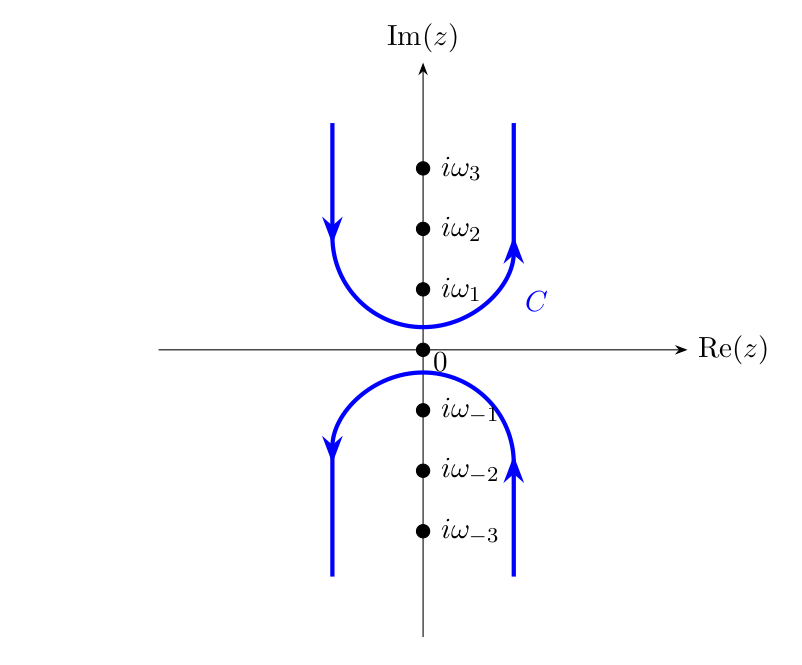}
        \caption{Integration contour $C$ containing the poles at $z=i\omega_n$ (Matsubara frequencies)}
    \label{fig:plot-b}
\end{figure}
Using this summation we convert the summation in $F^{(a)}_T$ as
\begin{equation}\label{a8}
 F^{(a)} =-\frac{a^2V}{24} \int \frac{d^3k}{(2\pi)^3} \oint_C \frac{dz}{2\pi i} \Big(\frac{ 4z^4 + 3k^4 - 6k^2z^2}{z^2 - k^2}\Big)\frac{1}{e^{\beta z}-1}
\end{equation}
Now we deform the contour as in Fig.(\ref{fig:plot-a}) which contains three simple poles $z=-k,0,+k$. We evaluate this using the residue theorem and obtain the expression given in Eq.(\ref{a9}).

\section {}
Re-writing the integral in Eq.(\ref{a13}) using $\frac{1}{e^{\beta x}-1}=\displaystyle\sum_{n=1}^{\infty}e^{-n\beta x}$ and using $\displaystyle\int_b^{\infty} d\bar{k}~\bar{k}^n e^{-\lambda \bar{k}}=\frac{e^{-\lambda b}}{\lambda}\displaystyle\sum_{i=0}^n\frac{n!}{i!}\frac{b^i}{\lambda^{n-i}}$, we evaluate the integral in Eq.(\ref{a13}) as
\begin{equation}\label{A1}
\begin{split}
    \mathcal{F}_T^{(a)} = &- \frac{a^2}{48\pi\beta^5} \sum_{m=1}^{\infty} \Bigg( \frac{24}{m^5} \bigg( \sum_{n=1}^{\infty}e^{-nm\pi\beta/L}\bigg) - \frac{24\beta}{m^5}\frac{\partial}{\partial\beta}\bigg( \sum_{n=1}^{\infty}e^{-nm\pi\beta/L}\bigg) + \frac{12\beta^2}{m^5}\frac{\partial^2}{\partial\beta^2}\bigg( \sum_{n=1}^{\infty}e^{-nm\pi\beta/L}\bigg) \\
    &- \frac{4\beta^3}{m^5}\frac{\partial^3}{\partial\beta^3}\bigg( \sum_{n=1}^{\infty}e^{-nm\pi\beta/L}\bigg) + \frac{\beta^4}{m^5}\frac{\partial^4}{\partial\beta^4}\bigg( \sum_{n=1}^{\infty}e^{-nm\pi\beta/L}\bigg) \Bigg)  
\end{split}    
\end{equation}
Using $\displaystyle\sum_{n=1}^{\infty}e^{-nx}=\frac{1}{2}\left(\coth{x}-1\right)$, we obtain the following
\begin{equation}\label{A2}
    \begin{split}
        \sum_{n=1}^{\infty}e^{-nm\pi\beta/L}&=\frac{1}{2}\coth\bigg(\frac{m\pi\beta}{2L}\bigg)-\frac{1}{2}\\
        \frac{\partial}{\partial\beta}\bigg( \sum_{n=1}^{\infty}e^{-nm\pi\beta/L}\bigg)&=-\frac{m\pi}{2L} \cosech^2\bigg(\frac{m\pi\beta}{2L}\bigg)\\
        \frac{\partial^2}{\partial\beta^2}\bigg( \sum_{n=1}^{\infty}e^{-nm\pi\beta/L}\bigg)&=\frac{m^2\pi^2}{2L^2} \coth\bigg(\frac{m\pi\beta}{2L}\bigg) \cosech^2\bigg(\frac{m\pi\beta}{2L}\bigg)\\
        \frac{\partial^3}{\partial\beta^3}\bigg( \sum_{n=1}^{\infty}e^{-nm\pi\beta/L}\bigg)&=-\frac{m^3\pi^3}{2L^3} \coth^2\bigg(\frac{m\pi\beta}{2L}\bigg) \cosech^2\bigg(\frac{m\pi\beta}{2L}\bigg)-\frac{m^3\pi^3}{4L^3} \cosech^4\bigg(\frac{m\pi\beta}{2L}\bigg)\\
        \frac{\partial^4}{\partial\beta^4}\bigg( \sum_{n=1}^{\infty}e^{-nm\pi\beta/L}\bigg)&=\frac{m^4\pi^4}{2L^4} \coth^3\bigg(\frac{m\pi\beta}{2L}\bigg) \cosech^2\bigg(\frac{m\pi\beta}{2L}\bigg)+\frac{m^4\pi^4}{L^4} \coth\bigg(\frac{m\pi\beta}{2L}\bigg)\cosech^4\bigg(\frac{m\pi\beta}{2L}\bigg)
    \end{split}
\end{equation}
Substituting Eq.(\ref{A2}) in Eq.(\ref{A1}), we obtain Eq.(\ref{a15}) 

\end{appendices}


\begin{thebibliography}{99}


\bibitem{CH-4-connes} A. Connes, \textit{Non-Commutative Geometry}, Academic Press, London (1994).

\bibitem{CH-4-dop} S. Doplicher, K. Fredenhagen, and J. E. Roberts, \textit{Phys. Lett. }{\bf B 331} (1994) 39; \textit{Commun. Math. Phys. }\textbf{172} (1995) 187.

\bibitem{hbg} H. B. G. Casimir, \textit{Koninkl. Ned. Akad. Wetenschap. Proc. }{\bf 51} (1948) 793.

\bibitem{pule} G. Plunien, B. Muller and W. Greiner, \textit{Phys. Rept. }{\bf 134} (1986) 87.

\bibitem{kam} K. A. Milton, \textit{The Casimir Effect Physical manifestations of Zero-Point Energy}, World Scientific, Singapore (2001).

\bibitem{b-m-m} M. Bordag, U. Mohideen and V.M. Mostepanenko, \textit{Phys. Rept. }\textbf{353} (2001) 1-205.

\bibitem{Mil} K. A. Milton, \textit{J. Phys. A: Math. and Gen. }{\bf 37} (2004) R209.

\bibitem{Brevik} I. Brevik, S. A. Ellingsen and K. A. Milton, \textit{New J. Phys. }{\bf 8} (2006) 236.

\bibitem{onofr} G. Bressi, G. Carugno, R. Onofrio and G. Ruoso, \textit{Phys. Rev. Lett. }{\bf 88} (2002) 041804.

\bibitem{brax} R. I. P. Sedmik and P. Brax, \textit{J. Phys: Conf. Ser. }(2018) 1138012014.

\bibitem{sedmik} R. I. P. Sedmik, \textit{Int. J. Mod. Phys. }{\bf A 35} (2020) 2040008; R. I. P. Sedmik and M. Pitschmann, \textit{Universe} {\bf 7} (2021) 234.

\bibitem{sed} U. Mohideen and A. Roy, \textit{Phys. Rev. Lett. }{\bf 81} (1998) 4549; V. M. Mostepanenko, \textit{J. Phys: Conf. Ser. }{\bf 161} (2009) 012003 and reference therein.

\bibitem{Bimonte} G. L. Klimchitskaya and V. M. Mostepanenko, \textit{Mod. Phys. Lett. }{\bf A 35} (2020) 2040007; G. Bimonte, B. Spreng, P. A. Maia Neto, G. L. Ingold, G. L. Klimchitskaya, V. M. Mostepanenko and R. S. Decca, \textit{Universe} {\bf 7} (2021) 93.

\bibitem{Wang} M. Wang, L. Tang, C. Y. Ng, R. Messina, B. Guizal, J. A. Crosse, M. Antezza, C. T. Chan and H. B. Chan, \textit{Nature Commun.} {\bf 12} (2021) 600.

\bibitem{oikonomou1} V. K. Oikonomou, \textit{Mod. Phys. Lett. }\textbf{A 24} (2009) 2405; \textit{Commun. Theor. Phys. }\textbf{55} (2011) 101.

\bibitem{CH-4-casadio} R. Casadio, A. Gruppuso, B. Harms and O. Micu, \textit{Phys. Rev. }{\bf D 76} (2007) 025016. 

\bibitem{CH-4-pinto} M. V. Cougo-Pinto, C. Farina and J. F. M. Mendes, \textit{Phys. Lett. }\textbf{B 529} (2002) 256.

\bibitem{kappa-casimir-korea} S. Nam, H. Park and Y. Seo, \textit{J. Korean Phys. Soc. }\textbf{42} (2003) 467.

\bibitem{CH-4-skp} E. Harikumar, S. K. Panja and V. Rajagopal, \textit{Nucl. Phys. }{\bf B 950} (2020) 114842.

\bibitem{kappa-spherical} H.-C. Kim, C. Rim and J. H. Yee, \textit{J. Korean Phys. Soc. }\textbf{53} (2008) 1826.

\bibitem{CH-4-skp3} E. Harikumar, K. V. Shajesh and S. K. Panja, \textit{Eur. Phys. J.} \textbf{C 84} (2024) 647.

\bibitem{CH-4-skp4} E. Harikumar, K. V. Shajesh and S. K. Panja, \textit{Ann. Phys. }\textbf{476} (2025) 169976.

\bibitem{snyder} S. A. Franchino-Vinas and S. Mignemi, \textit{Nucl. Phys. }\textbf{B 959} (2020) 115152.

\bibitem{CH-4-skp2} E. Harikumar, S. K. Panja, \textit{Int. Mod. Phys. Lett. }\textbf{A 38} (2023) 2350009.

\bibitem{CH-4-jerzy} F. Cianfrani, J. Kowalski-Glikman, D. Pranzetti and G. Rosati, \textit{Phys. Rev. }\textbf{D 94} (2016) 084044.

\bibitem{kappa-poincare} J. Lukierski, A. Nowicki, H. Ruegg and V. N. Tolstoy, \textit{Phys. Lett. }\textbf{B 264} (1991) 331; J. Lukierski, A. Nowicki and H. Ruegg, \textit{Phys. Lett. }\textbf{B 293} (1992) 344; J. Lukierski and H. Ruegg, \textit{Phys. Lett. }\textbf{B 329} (1994) 189; S. Majid and H. Ruegg, \textit{Phys. Lett. }\textbf{B 334} (1994) 348.

\bibitem{CH-4-hopf} S. Meljanac and M. Stojic, \textit{Eur. Phys. J. }\textbf{C 47} (2006) 531.

\bibitem{CH-4-mel2} S. Meljanac, S. Kresic-Juric and M. Stojic, \textit{Eur. Phys. J. }\textbf{C 51} (2007) 229.

\bibitem{CH-4-mel3} S. Meljanac, A. Samsarov, M. Stojic, and K. S. Gupta, \textit{Eur. Phys. J. }\textbf{C 53} (2008) 295.

\bibitem{Teo-1} S. C. Lim and L. P. Teo, \textit{J. Phys. A: Math. Theor. }\textbf{40} (2007) 11645.

\bibitem{geyer} B. Geyer, G. L. Klimchitskaya and V. M. Mostepanenko, \textit{Eur. Phys. J. }\textbf{C 57} (2008) 823.

\bibitem{Mota} V. B. Bezerra, G. L. Klimchitskaya, V. M. Mostepanenko and C. Romero, \textit{Phys. Rev. }\textbf{D 83} (2011) 104042; V. B. Bezerra, H. F. Mota and C. R. Muniz, \textit{Phys. Rev. }\textbf{D 89} (2014) 024015.

\bibitem{zhang} A. Zhang, \textit{Nucl. Phys. }\textbf{B 898} (2015) 220.

\bibitem{zhang1} A. Zhang, \textit{Phys. Lett. }\textbf{B 773} (2017) 125.

\bibitem{godel} A. F. Santos and F. C. Khanna, \textit{Phys. Lett. }\textbf{B 835} (2022) 137493.

\bibitem{matsubara1} T. Matsubara, \textit{Prog. Theor. Phys. }\textbf{14} (1955) 351.

\bibitem{matsubara2} L. Bellac, \textit{Thermal field theory}, Cambridge University Press (2000).

\bibitem{ok3} V. K. Oikonomou, \textit{J. Phys. }\textbf{A 40} (2007) 9929.

\bibitem{vishnu1} E. Harikumar and V. Rajagopal, \textit{Eur. Phys. J. }\textbf{C 79} (2019) 735.

\bibitem{bound} G. Bressi, G. Carugno, R. Onofrio and G. Ruoso, \textit{Phys. Rev. Lett. }\textbf{88} (2002) 041804.

\bibitem{panella} A. M. Frassino and O. Panella, \textit{Phys. Rev. }\textbf{D 85} (2012) 045030.

\bibitem{davis} P. C. W. Davies and P. Tee, \textit{Phys. Rev. }\textbf{D 110} (2024) 025009.

\bibitem{vishnu} V. Rajagopal, \textit{Gen. Rel. Grav. }\textbf{56} (2024) 14.

\bibitem{kappa-black} H.-C. Kim, C. Rim and J. H. Yee, \textit{Phys. Rev. }\textbf{D 76} (2007) 105012.

\bibitem{suman} D. Parai and S. K. Panja, \textit{Ann. Phys. }\textbf{470} (2024) 169815. 

\bibitem{ok4} V. K. Oikonomou, \textit{Int. J .Mod. Phys. }\textbf{A 25} (2010) 5935.

\bibitem{kappa-dirac} E. Harikumar, M. Sivakumar and N. Srinivas, \textit{Mod. Phys. Lett. }\textbf{A 26} (2011) 1103; E. Harikumar and V. Rajagopal, \textit{Int. J. Mod. Phys. }\textbf{A 35} (2020) 2050147.

\end{thebibliography}
\end{document}